# Ultrafast Photoluminescence from Graphene

Chun Hung Lui[1], Kin Fai Mak[1], Jie Shan[2], and Tony F. Heinz[1]*

[1]*Depts. of Physics and Electrical Engineering, Columbia University, 538 West 120th Street, New York, NY 10027, USA*

[2]*Dept. of Physics, Case Western Reserve University, 10900 Euclid Avenue, Cleveland, OH 44106, USA*

Since graphene has no band gap, photoluminescence is not expected from relaxed charge carriers. We have, however, observed significant light emission from graphene under excitation by ultrashort (30-fs) laser pulses. Light emission was found to occur across the visible spectral range (1.7 - 3.5 eV), with emitted photon energies exceeding that of the excitation laser (1.5 eV). The emission exhibits a nonlinear dependence on the laser fluence. In two-pulse correlation measurements of the time-domain response, a dominant relaxation time of tens of femtoseconds is observed. A two-temperature model describing the electrons and their interaction with strongly coupled optical phonons can account for the experimental observations.

* Corresponding author: tony.heinz@columbia.edu

The optical properties of graphene have attracted attention because of the insight they provide into the excited states of this remarkable material, and because of the potential that they offer for novel applications. Among the striking results is the absorbance of single-layer graphene of magnitude $\pi\alpha$, where $\alpha$ is the fine structure constant, in the near-infrared to visible spectral range [1, 2]. The possibility of tuning this absorption in the infrared by Pauli blocking has also been demonstrated [3, 4]. Optical measurements with ultrafast excitation pulses have provided means of probing electron and phonon dynamics in graphene [5-15]. To date, however, all investigations have been confined to probing the *light absorption* in graphene. Aside from the weak inelastic scattering associated with vibrations through the Raman process, there have been no reports of *light emission* from graphene. The lack of observable emission can be readily understood from the absence of a band gap in graphene. Carriers can fully relax through rapid electron-electron and electron-phonon interactions before the relatively slow process of light emission is possible. Thus, photoluminescence has only been reported in oxidized graphene [16], where the electronic structure has been modified and longer lived states may be present.

In this letter, we report the observation of significant light emission over a broad spectral range from pristine single-layer graphene under excitation by femtosecond laser pulses in the near infrared. This light emission process differs from conventional hot luminescence: it has a nonlinear dependence on the pump excitation and also appears at photon energies well above that of the excitation laser. We have characterized this emission process by measurements of the emission spectra and their dependence on pump fluence. We have also performed two-pulse correlation measurements of the emission process, which reveal a dominant response on the time scale of 10's of femtoseconds. These observations can be understood in a model in which the electronic excitations are largely thermalized among themselves, but are only partially equilibrated with strongly coupled optical phonons (SCOPs) and essentially decoupled from the other lattice vibrations. The femtosecond pump excitation can thus produce carriers with transient temperatures above 3000 K that give rise to readily observable emission in the visible range. In addition to revealing a new physical process in graphene, these measurements provide insight into carrier and phonon dynamics in graphene. The results indicate that electron-electron scattering under our experimental conditions is efficient on the 10-fs time scale, that coupling with the SCOPs is strong on a time scale below 100 fs, and that equilibration with other phonons occurs on a time scale approaching 1 ps.

In our experiment, we investigated single-layer graphene samples exfoliated from kish graphite (Toshiba) and deposited on freshly cleaved mica substrates. Information about the sample preparation and characterization is presented elsewhere [17]. The graphene samples were excited by ultrashort laser pulses with a photon energy of 1.5 eV from an 80-MHz modelocked Ti:sapphire oscillator. The pulse FWHM at the sample was 30 fs, as determined by a second-harmonic autocorrelation measurement. The spatial profile of the focused laser beam was characterized by scanning a sharp edge across the beam in the plane of the sample. The effective spot size was then determined by weighting this profile using the measured nonlinear fluence dependence of luminescence discussed below. The absorbed laser fluence $F$ was measured directly under the experimental excitation conditions [2]. It includes a modest absorption saturation effect observed at high fluences [10]. We measured the light emission under excitation both by individual pulses and by pairs of pulses. For the latter case, we recorded the light emission as a function of the temporal separation between two equivalent excitation pulses, which were orthogonally polarized to eliminate interference effects. The emission was collected in both transmission and reflection geometries and analyzed by a spectrometer coupled to a cooled charge-coupled device (CCD) array detector. The emission spectra were calibrated with a quartz tungsten halogen lamp. The emission strength is presented in terms of the spectral fluence $\mathcal{F}(\hbar\omega)$, i.e., total radiant energy emitted in all directions per unit area per unit photon energy as a function of the photon energy $\hbar\omega$. There is an estimated uncertainty of a factor of 10 in the absolute calibration of the emission strength. All measurements were performed under ambient conditions at room temperature.

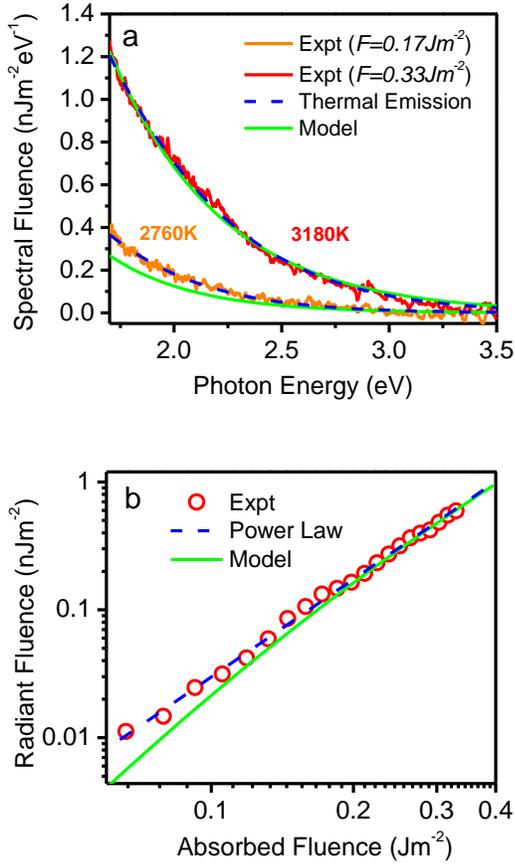

FIG. 1. (a) Spectral fluence of light emission from graphene for excitation with 30-fs pulses of absorbed fluences of $F = 0.17$ and $0.33$ Jm$^{-2}$. The spectra are compatible with the predictions for thermal emission (dashed blue lines), with $T_{em}$ = 2760 K and 3180 K, respectively. A full calculation using the two-temperature model described in the text also gives a good agreement (solid green lines). (b) A log-log plot of the measured total radiant fluence (red circles) for photons from 1.7 to 3.5 eV as a function of absorbed laser fluence. The data can be described by a power-law relation with an exponent of 2.5 (dashed blue line). The solid green line is a fit based on the two-temperature model. In both figures, the predictions of the model have been multiplied by a factor of 0.2 to match the scale of the experimental data.

Under excitation by femtosecond laser pulses, the graphene samples produced readily observable light emission over the entire spectral range from the visible to near-ultraviolet (1.7 - 3.5 eV). The emission was unpolarized and angularly broad. Two emission spectra for different absorbed laser fluences are shown in Fig. 1(a). Over the observed spectral range, the luminescence quantum efficiency was ~ $10^{-9}$. In contrast, for continuous-wave excitation of the same photon energy (1.5 eV), we could not detect any graphene light emission over the indicated spectral range (quantum efficiency < $10^{-12}$).

Another distinctive feature of the light emission process is its nonlinear dependence on the pump laser fluence. Fig. 1(b) displays the integrated radiant fluence over the observed spectral range (1.7 - 3.5 eV) as a function of the absorbed pump fluence. The emission varies with the absorbed fluence $F$ as a power law of $F^{2.5}$ [dashed blue line in Fig. 1(b)]. For light emission in different spectral windows we find a power-law relation, but with different exponents: an exponent of 2 for photons near the lower end of our spectral range and of 3.5 for photons at its upper end.

The experimental observations above immediately preclude several mechanisms for the light emission process. The emission of photons at energies above that of the pump photons and the nonlinearity of the process imply that we are not observing a conventional hot-luminescence process. Similarly, hot luminescence driven by a two-photon absorption process can also be excluded by the strong variation of the emission spectrum with pump fluence. The form of the observed emission spectra, however, provides a guide to the nature of the process. We see a steady decrease of light emission with increasing photon energy. This suggests comparison with the spectrum expected for thermal emission. For a system at an effective emission temperature $T_{em}$, we obtain from Planck's law, a spectral radiant fluence (integrated over all angles and polarizations) of

$$\mathcal{F}(\hbar\omega, T_{em}) = \tau_{em}\varepsilon(\hbar\omega)\frac{\omega^3}{2\pi^2 c^2}\left[\exp\left(\frac{\hbar\omega}{kT_{em}}\right)-1\right]^{-1}. \quad (1)$$

Here $\varepsilon(\hbar\omega)$ is the sample emissivity, which we determine directly from the measured absorption spectrum of graphene corrected for the influence of the mica substrate. Since we are describing the emitted energy, not the emitted power, the expression also contains a parameter $\tau_{em}$ to characterize the effective emission time for each laser excitation pulse.

This simple phenomenological description of the emission provides an excellent match to the experimental data [dashed blue curves in Fig. 1(a)]. The emission temperatures inferred from the shape of the spectra are, respectively, $T_{em}$ = 2760 K and 3180 K for absorbed fluences of 0.17 and 0.33 Jm$^{-2}$. The absolute magnitude of the experimental radiant fluence can be reproduced by $\tau_{em}$ in the range of 10 - 100 fs.

The analysis implies that carriers in graphene are well thermalized among themselves during the period of light emission. This finding suggests very rapid carrier-carrier scattering. The electrons and holes are initially created with a nearly monochromatic energy of 0.75 eV. During the period of light emission, which may occur on a time scale as short as that of the 30-fs excitation pulse, a largely thermalized energy distribution is apparently established for electrons and holes that contribute to the observed emission spectrum. This rapid thermalization is compatible with recent estimates of electron-electron scattering times [14, 18, 19]. For instance, for electron densities of ~$10^{12}$ cm$^{-2}$, the scattering times have been estimated to be tens of femtoseconds [18, 19]. Still shorter times would be expected under our experimental conditions with carrier excitation densities in the range of $10^{14}$ cm$^{-2}$.

The observed emission temperatures allow us to gain considerable insight into the emission mechanism. The emission temperature reflects the behavior of the electrons in the graphene, since they interact strongly with visible photons. Now if all absorbed laser energy were retained in the electronic system, the low electronic specific heat of graphene would lead to an electronic temperature of ~ 9000 K for the absorbed pump fluence of 0.33 Jm$^{-2}$. This is incompatible with the temperature of 3180 K extracted from the experimental emission spectrum. Therefore, even in this ultrafast light emission process, a significant fraction of the deposited energy must leave the electronic system. Since lateral diffusion of energy away from the excited region of the sample can be ruled out given the spatial dimensions and time scale, we conclude that efficient energy transfer to other degrees of freedom must occur. We note that in the limit of full equilibration of the excitation with *all* phonon degrees of freedom, *i.e.,* considering the full specific heat of graphene [20], we predict a temperature rise of only 380 K. Thus partial equilibration with the phonons must be considered.

The optical phonons in graphene serve as the most natural channel for energy relaxation from the excited electronic system, since electrons in graphene are strongly coupled to optical phonons near the Γ and K points in the Brillouin zone [12, 21]. Investigations of phonon dynamics in graphite and carbon nanotubes by time-resolved Raman spectroscopy have directly demonstrated energy transfer from photoexcited electrons to these strongly coupled optical phonons (SCOPs) within 200 fs [22, 23]. Various theoretical and experimental studies have also obtained ultrashort (< 100 fs) emission times for optical phonons in graphene [18, 24], graphite [13, 15, 25], and carbon nanotubes [26, 27].

To analyze the results further, we introduce a model of excitations in the electronic system and in the SCOPs, each characterized by its respective temperature, $T_{el}$ and $T_{op}$, and linked by the electron-phonon coupling:

$$\frac{dT_{el}(t)}{dt} = \frac{I(t) - \Gamma(T_{el}, T_{op})}{c_e(T_{el})},$$

$$\frac{dT_{op}(t)}{dt} = \frac{\Gamma(T_{el}, T_{op})}{c_{op}(T_{op})} - \frac{T_{op}(t) - T_0}{\tau_{op}}. \qquad (2)$$

In this description, the graphene is excited by the absorbed irradiance $I(t)$, which initially excites the electronic system. Energy then flows into the SCOPs at a rate described by $\Gamma(T_{el}, T_{op})$. This quantity is constructed based on consideration of available phase space for scattering of the excited electrons and includes only one adjustable parameter to characterize the overall rate. The specific heat of the electrons (per unit area) is denoted by $c_{el}$, while that of the SCOPs is $c_{op}$. These quantities are obtained, respectively, from theory and experimental data using Raman spectroscopy to determine phonon populations under femtosecond laser excitation. In addition to energy flow between the electrons and the SCOPs, we have included a slower coupling of the SCOPs to other phonons in the system through anharmonic decay. This channel for energy flow is described simply by a relaxation time $\tau_{op}$, which is estimated from experimental measurements of time-resolved Raman scattering in related systems [22, 23]. We neglect the heating of these more numerous secondary phonons and assume that they remain at the ambient temperature of $T_0$ = 300 K. A detailed description of the parameters in the model is presented in the supplemental material [28].

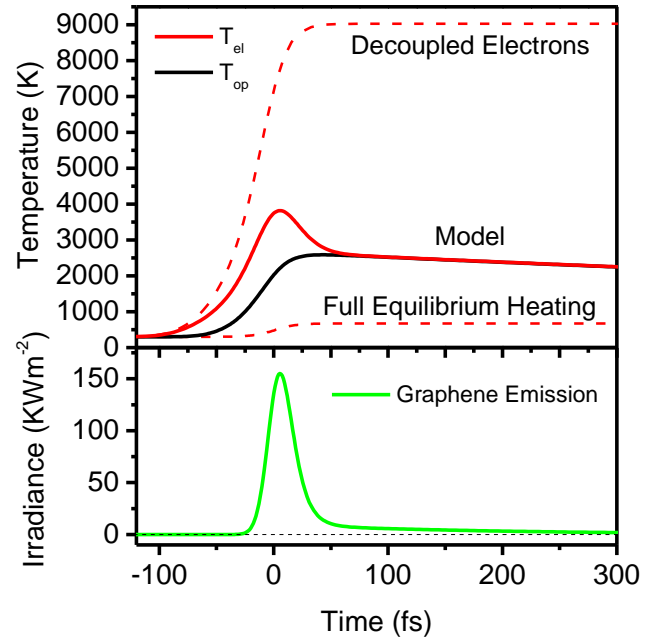

FIG. 2. Simulations using the two-temperature model (described in the text) of the temporal evolution of the electronic temperature $T_{el}$ (red line), the SCOPs temperature $T_{op}$ (black line), and of the resulting graphene light emission (green line) for photon energies from 1.7 to 3.5 eV. The absorbed fluence $F$ of the 30-fs pump pulse is 0.33 Jm$^{-2}$. For comparison, the upper panel also shows the calculated electronic temperatures for completely decoupled electrons and for full equilibrium of all degrees of freedom of the graphene sample (dashed red lines).

Fig. 2 displays the predicted temporal evolution for the temperatures of electrons ($T_{el}$) and SCOPs ($T_{op}$), as well as the corresponding light emission from graphene, for our experimental conditions. For comparison, we also show the electronic temperatures for the completely decoupled electronic system and for full thermal equilibrium of the graphene sample. These limits are, as discussed above, clearly incompatible with the experimental results. Within the two-temperature model, rapid energy transfer from electrons to SCOPs occurs during the laser excitation process. This results in a significant decrease in the electronic temperature compared to the case of uncoupled electrons (from a peak of ~ 9000 K to ~ 3800 K for $F = 0.33$ Jm$^{-2}$). Equilibration with the SCOPs is almost complete within 50 fs, with the electronic system having lost over 95% of its energy to the SCOPs. Using the calculated $T_{el}$, we find from Eqn. 1 both the predicted emission spectrum and the integrated light emission. The results agree well with the experimental spectra and fluence dependence, as shown, respectively, in Figs. 1(a) and (b). We also note that the effective emission temperature $T_{em}$ inferred from the experimental time-integrated emission spectra approaches the peak value of the electronic temperature. This is due to the strong nonlinear dependence of the emission on the electron temperature [28].

To probe the dynamics of the light emission process more directly in the time domain, we performed two-pulse correlation measurements in which the total radiant fluence (over photon energies in the range of 1.7 - 3.5 eV) was measured as a function of the temporal separation between a pair of laser excitation pulses. Fig. 3 shows the resulting correlation trace for an absorbed fluence of $F = 0.17$ Jm$^{-2}$. A dominant response on the time scale of 10's fs is observed, with weaker, slower decay extending over 100's fs. The form of the correlation trace, with its dominant short response time, is seen under all conditions. The details, however, vary with the spectral range of the detected photons, as well as with the pump fluence. If we restrict detection to the high-energy photons, for example, we observe a shorter response time than that obtained by detecting only the low-energy photons [28]. This effect can be understood as a consequence of the dependence of the emission strength on the electronic temperature for different photon energies, *i.e.,* the relation is more nonlinear for higher photon energies than for lower photon energies.

We have applied the two-temperature model presented above to analyze the two-pulse correlation data. The underlying origin of the correlation feature can be understood from the calculation of the electronic temperature under two-pulse excitation (Fig. 4). When the two pulses are sufficiently close to one another, the peak electronic temperature achieved by the second pulse exceeds that from one pulse alone. Since the light emission process is strongly nonlinear in temperature, we then observe a greater signal than for the two fully separated pulses. The enhancement is strongest at very short pulse separations, where electrons remain partially out of equilibrium with the SCOPs. Aweaker enhancement of the emission persists during the slower decay of the subsystems of equilibrated electrons and SCOPs. Carrying out full calculation within the model yields good agreement with the measured two-pulse correlation function (green line in Fig. 3).

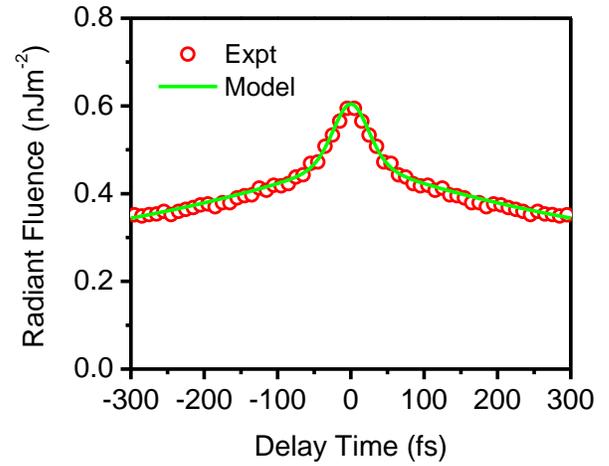

FIG. 3. Total radiant fluence emitted by graphene over photon energies between 1.7 and 3.5 eV as a function of temporal separation between two identical laser excitation pulses. The absorbed fluence $F$ of each pulse is 0.17 Jm$^{-2}$. The data for positive and negative delays were averaged to increase the signal-to-noise ratio. The red circles are the experimental data; the green line is the prediction of the two-temperature model, multiplied by a factor of 0.2 to match the magnitude of the experimental data.

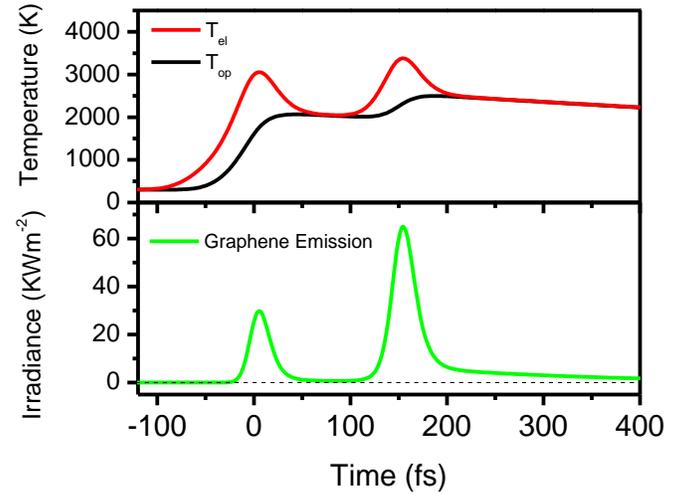

FIG. 4. Simulations as in Fig. 2, but with excitation by a pair of laser pulses, each yielding an absorbed fluence of $F = 0.17$ Jm$^{-2}$ and separated in time by 150 fs.

In conclusion, the observed spectrally broad light emission from graphene can be understood as a direct consequence of a transient regime in which the electron distribution is driven strongly out of equilibrium with the phonons by ultrafast laser excitation. The existence of such energetic electron distributions has been reported in many different condensed-matter systems. Our work suggests that light emission may also be observable from such hot electrons in these materials. Characterization of spectra and dynamics of light emission would then provide a new and valuable window to probing electron dynamics.

The authors would like to thank D. Boschetto, K. Ishioka, P. V. Klimov, Z. Q. Li and H. G. Yan for helpful discussions. The authors acknowledge support from the National Science Foundation under grant CHE-0117752 at Columbia and grant DMR-0907477 at Case Western Reserve University; from the Air Force Office of Scientific Research under the MURI program; and from the New York State Office of Science, Technology, and Academic Research (NYSTAR).

----------------------------------

## Supplemental Material

1. **The two-temperature model**

Here we discuss the details of the two-temperature model, as embodied in Eqn. 2 of the main text, for the temporal evolution of the electronic temperature $T_{el}$ and of the temperature $T_{op}$ of the strongly coupled optical phonons (SCOPs). In the two-temperature model these quantities are determined by the absorbed laser irradiance $I(t)$, the electronic specific heat $c_{el}$, the specific heat $c_{op}$ for the SCOPs, the lifetime $\tau_{op}$ of the SCOPs arising from coupling to other phonons, and the electron-SCOP energy exchange rate $\Gamma(T_{el}, T_{op})$. We consider each of these quantities below.

We model the temporal profile of the ultrafast excitation pulse using the form $I(t) = (F/2\tau_{exc})\operatorname{sech}^2(t/\tau_{exc})$, where $F$ denotes the absorbed fluence and $\tau_{exc}$ the duration of the exciting laser pulse. The pulse duration was determined by a second-harmonic autocorrelation measurement and yielded a value of $\tau_{exc}$ = 19 fs. The absorbed fluence was established by measurement of the absorbed energy combined with a determination of the spatial profile of the beam.

For the electronic specific heat $c_{el}$ (per unit area), we used the following analytical result derived from the linear dispersion of the graphene bands:

$$c_{el}(T_{el}) = \frac{18\zeta(3)}{\pi(\hbar v_F)^2} k^3 T_{el}^2 \quad . \tag{S1}$$

Here $\zeta(3)$ = 1.202 is the zeta function, $v_F = 1.1 \times 10^6 \, ms^{-1}$ is the Fermi velocity of electrons in graphene, and $k$ is the Boltzmann constant.

The electrons in graphene are strongly coupled to only a small fraction of optical phonons in the Brillouin zone near the $\Gamma$ and K points; these are the strongly coupled optical phonons (SCOPs). The SCOPs scatter electrons from one part of the Dirac cone to another (Γ point phonons) or between the Dirac cones (K point phonons). In our model, we assumed that only these SCOPs are directly excited by the electrons through the electron-phonon interaction. The coupling of these SCOPs to other phonon modes is described by a phenomenological parameter, the lifetime of the SCOPs $\tau_{op}$. In our analysis, we used $\tau_{op}$ = 1.5 ps, based on the measured anharmonic decay rates in carbon nanotubes (1.1ps) [S1] and graphite (2.2ps) [S2]. Further, for simplicity, we neglected the (weak) dispersion in the phonon energy and considered all phonons to have the same energy of 200 meV as for the Γ-point phonon.

The specific heat $c_{op}$ (per area) of the SCOPs in graphene was determined from the time-resolved Raman studies of graphite [S2]. These measurements yielded the population of the SCOPs (before any significant anharmonic decay occurred) as a function of the deposited laser excitation density per graphene layer. The SCOP temperature extracted from the phonon population is found to increase sublinearly with the excitation power, *i.e.*, the specific heat increases nonlinearly with temperature. Fitting these data provides an expression for the specific heat of the SCOPs as a function of $T_{op}$ in the temperature range between 500 K and 2500 K as follows:

$$c_{op}(T_{op}) = -4.79 \times 10^9 + 9.09 \times 10^6 T_{op} + 4453 T_{op}^2 + 1.29 T_{op}^3 \qquad (S2)$$

(in units of eVcm$^{-2}$K$^{-1}$).

We construct the electron-SCOP energy exchange rate $\Gamma(T_{el}, T_{op})$ using the available phase space for scattering of electrons near the K-point of the Brillouin zone. The complete expression of $\Gamma(T_{el}, T_{op})$ includes both the emission of a phonon (the first term) and the absorption of a phonon (the second term):

$$\Gamma(T_{el}, T_{op}) = \alpha \left\{ [1 + n(T_{op})] \int D(E) D(E - \hbar\Omega) f(E, T_{el}) [1 - f(E - \hbar\Omega, T_{el})] dE \right.$$
$$\left. - n(T_{op}) \int D(E) D(E + \hbar\Omega) f(E, T_{el}) [1 - f(E + \hbar\Omega, T_{el})] dE \right\} \quad . \quad \text{(S3)}$$

Here $D(E) = (2E/\pi)(\hbar v_F)^2$ is the density of states of electrons in graphene, including the spin and valley (*i.e.* K and K' points) degeneracies. The term $n(T_{op}) = [\exp(\hbar\Omega/kT_{op}) - 1]^{-1}$ represents the SCOP population at temperature $T_{op}$, and $f(E, T_{el}) = [\exp(E/kT_{el}) + 1]^{-1}$ is the Fermi-Dirac distribution for electrons at electronic temperature $T_{el}$. In the later expression, we assume that the electrons and holes are in equilibrium with one another and have zero chemical potential. The only adjustable parameter in our model then is the proportionality constant *α* that represents the overall electron-phonon coupling strength. Throughout the analysis in this work, we have used the value of α = 5 eV$^2$cm$^2$s$^{-1}$ that represents the best fit to all the data.

We note that in our model, we neglected decay channels for the electronic excitation involving the substrate. Electron scattering by substrate phonons has been invoked to explain transport data in graphene [S3]. However, under excitation by high intensity ultrafast laser excitation, similar behaviour for light emission was observed for graphene on different (silicon dioxide and mica) substrates, as well as from bulk graphite. This suggests that substrate coupling plays a minor role in the relevant material response for our measurements.

2. **Role of time integration on the form of the emission spectra**

In the experimental measurements, the recorded spectra for light emission from graphene were integrated over time. Within our description of the emission process, these time-integrated spectra thus correspond to light emission occurring at differing electronic temperatures of the excited graphene. However, over the observed spectral range (see Fig. 1(a) of the main text), the experimental data are found to be described quite well by thermal emission spectra at a single effective temperature. Here we discuss the origin of this behavior. The essential factor is that we are probing only the high-energy tail of the emission spectrum, which increases strongly with increasing temperature. This causes the integrated emission spectrum to weight predominantly the behavior near the peak temperature.

To examine this behavior in more detail, we consider the predicted emission for the temporal evolution of the electronic temperature $T_{el}(t)$ in Fig. 2 of the main text, as derived from the two-temperature model for our experimental parameters. We present this result over a longer time range in Fig. S1(a). We see that $T_{el}$ reaches the peak value of 3800 K and drops to 2700 K within 50 fs. This rapid response corresponds to the electronic system remaining out of equilibrium with the SCOPs. Once equilibrium is established between these two subsystems, the temperature falls below 1000 K on the time scale of a few picoseconds. Using the expression for thermal emission from graphene (Eqn. 1 of the main text), we calculated the expected spectra for photon energies in the range of 1.75 - 3.5 eV integrated or different time intervals, as shown in Fig. S1(b). We see that the emission for this range of photon energies arises primarily from emission near the peak electronic temperature. As can be seen in the figure, emission occurring after 50 fs changes the spectrum only modestly. Further, for times greater than 400 fs, hardly any emission is expected for the given spectral range. Since the range of temperatures for which strong emission occurs is relatively limited, we anticipate that a fit of the integrated spectrum to that of an effective emission temperature (by Eqn. 1 of the main text) will work rather well. This is shown to be the case in Fig. S1(b), which yields effective temperatures of 3550 K and 3150 K, respectively, for the spectra obtained for emission over 50 fs and 10 ps intervals.

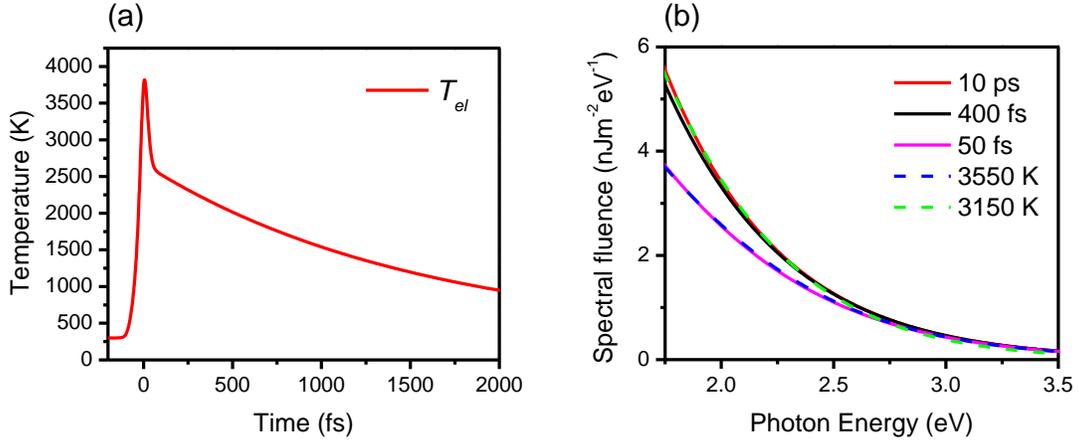

Fig. S1: (a) The temporal evolution of the electronic temperature $T_{el}(t)$ obtained from the two-temperature model for our experimental parameters as described in the main text. (b) Integrated emission spectra calculated for the electronic temperature profile $T_{el}(t)$ of (a) over times from -100 fs to 50 fs, 400 fs, and 10 ps. The integrated spectra at 50 fs and 10 ps are described well by thermal emission spectra, respectively, at constant effective temperatures of 3550 K and 3150 K, as shown.

## 3. Emission Temperature as a function of the absorbed laser fluence

In the main text we examined experimental data for the overall emission strength as a function of the absorbed laser fluence. Additional information about the graphene response can be obtained by considering the variation of the effective emission temperature as a function of the absorbed laser fluence.

The measured effective emission temperatures are shown in Fig. S2 (symbols) for various values of pump fluence. The experimental results, determined by fitting the spectra to the thermal emission form [Eqn. (1) of the main text], are seen to compare well with the predictions of the two-temperature model (Fig. S2, solid line). In applying the two-temperature model, the effective emission temperatures were obtained from a fit of the time-integrated spectra as described in Sect. 2. The observed sublinear increase of the electronic temperature with the pump fluence is a manifestation of the quadratic temperature dependence of the electronic specific heat, as well as of the strong dependence of the electron-phonon coupling on the electronic temperature.

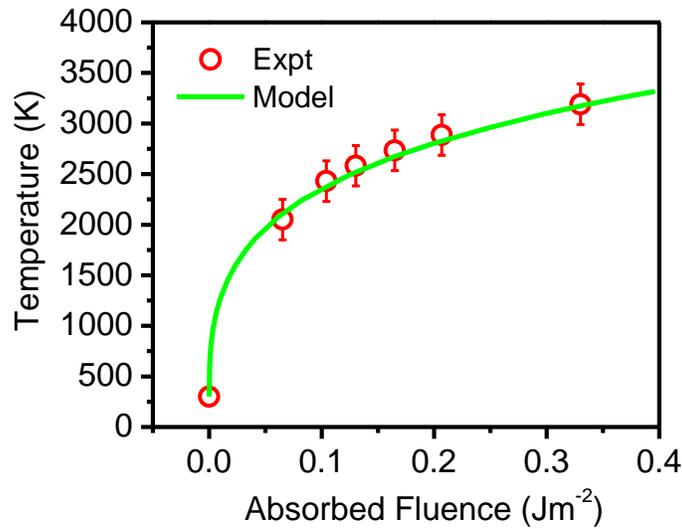

Fig. S2: The effective emission temperature of graphene as a function of absorbed fluence of the pump laser. The red circles are obtained from the experimental emission spectra by fitting to the thermal emission formula. The error bars represent the uncertainty in fitting the emission spectra by the thermal emission formula. The data point at zero fluence corresponds to room temperature (300 K). The green line is the predicted behavior within the two-temperature model.

## 4. Dependence of two-pulse correlation data on the measured emission spectral range

For the two-pulse correlation measurements, we observed different correlation traces when recording signals for light emission over different ranges of photon energy. As shown in Fig. S3(a), if we restrict detection to low-energy photons (1.75 – 2 eV), we observe a weaker enhancement when the two pulses overlap and a longer response time than that obtained by detecting high-energy photons (2.5 - 2.75 eV, Fig, S3(b)). This behavior can be well explained by the two-temperature model (green lines in Fig. S3). It reflects the dependence of the emission strength on the electronic temperature for different photon energies. The relation is more strongly nonlinear for higher photon energies than for lower photon energies. Thus, a given underlying variation in the peak electronic temperature with temporal separation of the two excitation pulses yields somewhat different shapes in the correlation feature.

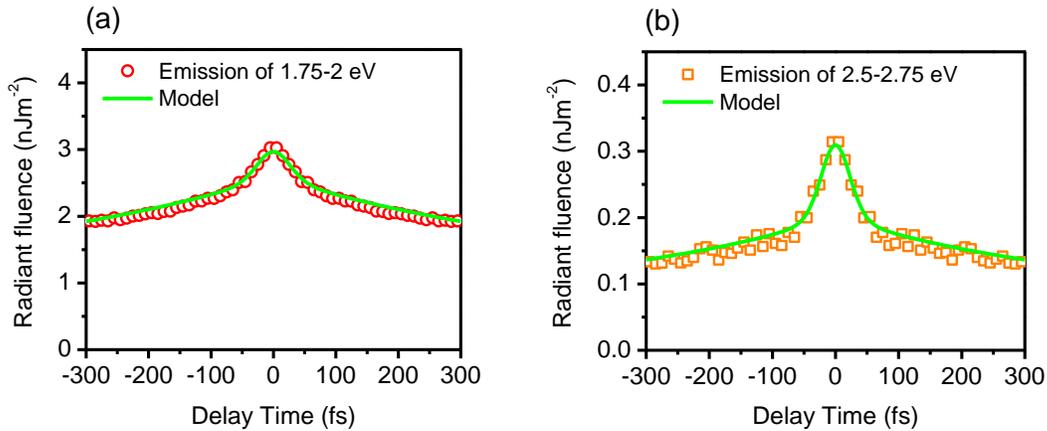

Fig. S3: Total radiant fluence emitted by graphene over photon energies between 1.7 and 2 eV (a) and between 2.5 and 2.75 eV (b) as a function of the temporal separation between two identical laser excitation pulses. The absorbed fluence $F$ of each pulse is 0.17 Jm$^{-2}$. The data for the positive and negative delays were averaged to increase the signal-to-noise ratio. The symbols represent the experimental data and the green lines are the predictions of the two-temperature model. The results from the model have been multiplied by a factor of ~ 0.2 to match the emission strength measured in the experiment.